\documentclass[aps,prl,twocolumn,superscriptaddress,floatfix]{revtex4-1}
 
\usepackage{graphicx}% Include figure files
\usepackage{dcolumn}% Align table columns on decimal point
\usepackage{verbatim}
\usepackage[colorlinks=true]{hyperref}% add hypertext capabilities
\usepackage{multirow}
\usepackage{bm}% bold math
\usepackage{times}
\usepackage{color}
\usepackage{amstext}
\usepackage{amsmath}
\usepackage{amssymb} 
\usepackage{amsfonts}
\usepackage{natbib}
\usepackage[final]{microtype}
\usepackage{csquotes,lmodern}
\usepackage{hyperref}
\hypersetup{hidelinks}
\newcommand{\ci}[1]{$_{\text{#1}}$} %Lower index in chemical formulae - e.g. Bi\ci{2}Se\ci{3} gives formulae for Bi2Se3
\newcommand{\uc}{u.c.{\ }}

\begin{document} 

\title{Approaching two-dimensional superconductivity in ultrathin DyBa\ci{2}Cu\ci{3}O\ci{7-$\delta$}}

\author{R.~D.~Dawson}
\affiliation{
Max Planck Institute for Solid State Research, Heisenbergstra{\ss}e~1,
70569 Stuttgart, Germany}

\author{K.~S.~Rabinovich}
\affiliation{
Max Planck Institute for Solid State Research, Heisenbergstra{\ss}e~1,
70569 Stuttgart, Germany}

\author{D.~Putzky}
\affiliation{
Max Planck Institute for Solid State Research, Heisenbergstra{\ss}e~1,
70569 Stuttgart, Germany}

\author{G.~Christiani}
\affiliation{
Max Planck Institute for Solid State Research, Heisenbergstra{\ss}e~1,
70569 Stuttgart, Germany}

\author{G.~Logvenov}
\affiliation{
Max Planck Institute for Solid State Research, Heisenbergstra{\ss}e~1,
70569 Stuttgart, Germany}

\author{B.~Keimer}
\affiliation{
Max Planck Institute for Solid State Research, Heisenbergstra{\ss}e~1,
70569 Stuttgart, Germany}

\author{A.~V.~Boris} 
\email{A.Boris@fkf.mpg.de}
\affiliation{
Max Planck Institute for Solid State Research, Heisenbergstra{\ss}e~1,
70569 Stuttgart, Germany}

%\date{\today}

        \begin{abstract}
        
        The temperature dependence of the superfluid density $\rho_s(T)$ has been measured for a series of ultrathin MBE-grown DyBa\ci2Cu\ci3O\ci{7-$\delta$} superconducting (SC) films by sub-mm wave interferometry combined with time-domain THz spectroscopy and IR ellipsometry. We find that all films 10 \uc and thicker show the same universal temperature dependence of $\rho_s(T)$, which  follows the critical behavior characteristic of single crystal YBa\ci2Cu\ci3O\ci{7-$\delta$} as $T$ approaches $T_c$. In 7 u.c. thick films, $\rho_s(T)$ declines steeply upon approaching $T_c$, as expected for the Berezinskii-Kosterlitz-Thouless vortex unbinding transition. Our analysis provides evidence for a sharply defined 4 \uc non-SC interfacial layer, leaving a quasi-2D SC layer on top. We propose that the SC state in this interfacial layer is suppressed by competing (possibly charge) order.
\end{abstract}


\maketitle
The commonality of the layered CuO$_2$ plane structure to all families of copper oxide high-$T_c$ superconductors (HTSCs) implies that understanding the effect of reduced dimensionality on superconductivity in the cuprates is key to elucidating the mechanism of HTSC. Even in the presence of strong anisotropy, the CuO$_2$-bilayer material $R$Ba\ci2Cu\ci3O\ci{7-$\delta$} ($R$BCO), where $R$ = rare earth, displays critical fluctuations belonging to the 3D-XY universality class, where the fluctuation correlations extend over many CuO\ci2 planes along the $c$-axis \cite{BonnPRL1994,Meingast2001,Grbic2011}. Close to optimal doping the 2D-like properties of the SC condensate appear only within a very narrow temperature range just below $T_c$ \cite{Nizhankovskiy2019},where uncorrelated fluctuations in adjacent CuO$_2$ layers are described by the Berezinskii-Kosterlitz-Thouless (BKT) vortex unbinding transition \cite{Kosterlitz2016,Minnhagen1987}.

One way to make the manifestation of 2D-SC noticeable is to reduce the coupling between the CuO\ci2 planes by doping. Indeed in the underdoped regime, where adjacent CuO$_2$ bilayers are weakly Josephson coupled, the SC fluctuations persist over a wider temperature range both above and below $T_c$ \cite{Meingast2001,Grbic2011,Cyr2018}. Furthermore, intrabilayer SC precursor fluctuations may survive up to temperatures much higher than the maximum $T_c$ due to coherent tunneling of preformed Cooper pairs within the CuO\ci2 bilayer units \cite{Dubroka2011}. Despite this, no evidence has yet been reported for the BKT transition in such extremely anisotropic HTSC systems \cite{Broun2007}. In this case the characteristic 2D features are obscured by the complex relationship between 2D and 3D fluctuations caused by Josephson coupling between the planes \cite{Blatter1994} and a variety of competing orders \cite{Keimer2015,Cyr2018}.

Another route to approach the 3D to 2D crossover is to reduce the sample thickness to less than the $c$-axis correlation length. Early reports of the BKT transition based on $dc$ transport measurements in ultrathin YBCO layers sandwiched between semiconducting (Pr,Y)BCO are controversial \cite{Matsuda1993, Repaci1996,Kosterlitz2016}. Vortex diffusion in such layers has been studied by single-coil mutual inductance (MI) measurements at microwave frequencies (MHz to GHz) and discussed in terms of the dynamical BKT theory, although vortex-antivortex pairing is strongly obscured by defect pinning \cite{Gasparov2005}. More recently, the BKT-like downturn of superfluid density $\rho_s$ with temperature has been reported from low frequency (kHz) two-coil MI measurements of 2 u.c. thick highly reduced Ca-doped YBCO \cite{Hetel2007}. To date, however, similar behavior in optimally doped $R$BCO films has not been reported.

In this Letter, we report the temperature and thickness dependence of $\rho_s$ in a series of near-optimally doped \mbox{DyBCO} ultrathin films grown by molecular beam epitaxy (MBE). For films thicker than 10 \uc we observe a universal temperature dependence of $\rho_s(T)$, while for thinner films signatures of the BKT transition emerge and reveal 2D-SC fluctuations over an exceptionally broad temperature range of $\sim$  13 K above $T_c$. Our results also provide evidence for a $\sim$ 4 \uc non-SC layer at the substrate interface that shares a sharp boundary with the SC portion of the film, and we propose a picture wherein an epitaxially stabilized competing order suppresses the SC in the interfacial layer while leaving the CuO\ci2 planes above superconducting.

DyBCO films with thicknesses ranging from 7 to 60 \uc (8 - 70 nm) were grown on (100)-oriented (LaAlO\ci3)\ci{0.3}(Sr\ci2AlTaO\ci6)\ci{0.7} (LSAT) substrates of dimensions 10 x 10 mm$^2$ by ozone-assisted atomic-layer-by-layer oxide MBE, and the high crystal structure quality of the films was confirmed by X-ray diffraction and transmission electron microscopy measurements, as reported elsewhere \cite{Putzky2020}. The results reported here are primarily based on measurements of the complex transmission obtained with a tabletop quasi-optical submillimeter-wave Mach-Zehnder interferometer. These measurements were supplemented by broadband measurements of the complex dielectric function via combined time-domain THz spectroscopy and far-infrared-to-UV ellipsometry \cite{SM}. 
                \begin{figure}
                    \includegraphics[width=76mm]{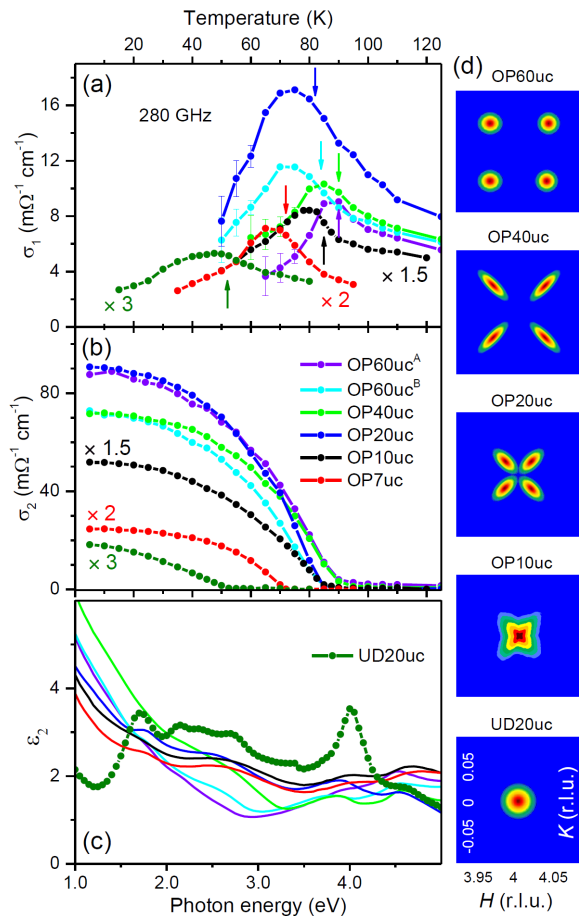}
                    \caption{(a),(b) The temperature dependence of the real
and imaginary parts of the optical conductivity, $\tilde{\sigma} = \sigma_1 + i \sigma_2$, measured at 280 GHz for all samples studied in this work. The arrows indicate the values of $T_c$. (c) Imaginary part of the dielectric function for DyBCO films in the visible-near-UV spectral range, as obtained by ellipsometry at $T$~=~293~K. (d) Illustrations of high-resolution
XRD reciprocal space maps for the $HK$-plane centered about the (4 0 10) reflection adapted from Ref.\cite{Putzky2020}. \label{AllDyBCO}}
                \end{figure}
                
Figure 1 shows the (a) real and (b) imaginary parts of the complex conductivity, $\tilde{\sigma} = \sigma_1 + i \sigma_2$, extracted from the measured transmission and phase shift of the film-on-substrate systems at 280 GHz. The imaginary part of the conductivity increases sharply below $T_c$ due to the increase in screening caused by the formation of the SC condensate. The real part of the conductivity shows a maximum below $T_c$ which corresponds to the increase in the scattering time, $\tau(T)$, of thermally excited quasiparticles. %The increasing error bar in $\sigma_1$ with decreasing temperatures reflects %the relatively large increasing contribution from the imaginary part that %masks the absorptive response, particularly for thick films with large superfluid %density.
To determine the doping level in the films, we performed spectroscopic ellipsometry measurements in the visible range, shown in Fig. 1(c). All samples with thicknesses equal to and greater than 10 \uc and $T_c$ values ranging from 80 to 90 K are nearly optimally doped, as indicated by the absence of a sharp peak in the imaginary part of the dielectric function at 4 eV. Such a peak, which is specific to the 123 family of HTSCs in contrast to other cuprates, is attributed to the dumbbell Cu 3$d_{3z^2-1}$ to 4$p_x$ transitions and allows us to resolve the presence of intact CuO chains \cite{Kircher1989, Cooper1992}. For comparison, we also show the response of a 20 \uc film (UD20uc) annealed at lower ozone pressure with reduced $T_c$ = 52 K, which clearly shows the presence of the 4 eV peak. Interestingly, a 7 \uc film (OP7uc), which has a reduced $T_c$ of 72 K compared to that of 85 K for a similar 10 \uc film (OP10uc), reproducible even after annealing and reoxidizing, displays very similar optical spectra to OP10uc, implying nearly the same doping in those samples.

Strain effects in the films were investigated by high-resolution X-ray diffraction measurements of the $HK$-plane centered about the (4 0 10) reflection, depicted in Fig. 1(d) as illustrations adapted from Ref.\cite{Putzky2020}. The results show that for all near-optimally doped films the structure is dominated by four-fold orthorhombic twin domains. The 60 \uc films have a relaxed orthorhombic structure as indicated by the four circular diffraction spots in the upper reciprocal space map. The four-fold diffraction pattern displays a thickness dependent elongation and shift toward the center, tracking the evolution of the decreasing strain relaxation with decreasing film thickness. In contrast to the single circular diffraction spot for an underdoped film, which suggests a tetragonal structure, the thinnest samples display a propeller-like diffraction pattern, indicating that they adapt to the square structure of the underlying substrate lattice while also retaining CuO chains.

From the measured complex conductivity the penetration depth $\lambda$, which corresponds to the superfluid density $\rho_s = 1/\lambda^{2}$, can be obtained as $\lambda^{-2}(T) = \lim_{\omega\to0} \mu_0 \omega \sigma_2(T)$. The frequency used in Fig. 1(a,b) is within the London limit, where $\sigma_2 \propto \omega^{-1}$, so the measured $\sigma_2$ provides a close estimate of $\lambda$. Detailed analysis requires the frequency and temperature dependence of the complex conductivity, $\tilde{\sigma}(\omega, T)$, which we measured over a broad range of frequencies (0.1 meV $<\hbar \omega<$ 1 eV) and temperatures (5 K $< T <$ 300 K) \cite{SM}. The in-plane superfluid density $\rho_s(T)$ was accurately extracted by fitting the spectra with the two-fluid model, $\tilde{\sigma}(\omega,T) = i \rho_n(T)/(\omega + i/\tau) + i \rho_s(T)/\omega$, with $\rho_n + \rho_s$ remaining constant. Our analysis shows that changes in $\sigma_2$ are quantitatively consistent with the opening of the SC gap in $\sigma_1$ and its spectral weight transfer into the delta function at $\omega = 0$. Kramers-Kronig consistency of the measured $\sigma_1(\omega,T)$ and $\sigma_2(\omega,T)$ up to near-IR frequencies imply that the Ferrell-Glover-Tinkham sum rule holds in DyBCO to within 2\% error \cite{SM}. This analysis shows that the two-fluid model is obeyed, implying in particular that the temperature dependence of $\tau(T)$ is responsible for the peak in $\sigma_1(T)$ below $T_c$ \cite{Bonn1999,Fink2000}.
The different shape of the peaks in Fig. 1(a) implies the presence of different interfacial scattering and impurity levels due to different growing conditions of the films. 

Figure 2(a) shows the normalized values $\rho_s(T)/\rho_{s0}$ for the near-optimally doped films with thicknesses 10 to 60 \uc Strikingly, the evolution of the superfluid density with temperature is the same for all samples, implying the universality of $\rho_s(T)$ for the near-optimally doped DyBCO films, in spite of their different impurity and strain relaxation levels. 
We compare this temperature dependence with that for ultra-clean single crystal YBCO, shown by the solid gray line in Fig. 2(a) \cite{Bonn1998}, averaged over $a$ and $b$ crystallographic directions of a detwinned crystal. Near $T_c$ our data are reasonably consistent with the single crystal data and show the same critical behavior, which  has been taken as evidence for 3D-XY critical fluctuations over a broad temperature range \cite{BonnPRL1994}. Below $0.8\cdot T_c$ our data deviates from the single crystal data. Linear dependence of $\rho_s(T)$ at low temperatures has been a key piece of evidence for nodes in the energy gap in cuprate HTSCs \cite{Hardy1993}. In contrast, our data approaches a $T^2$ dependence, characteristic of nodal pair breaking by disorder \cite{LeeHone2017}.

                \begin{figure}
                    \includegraphics[width=76mm]{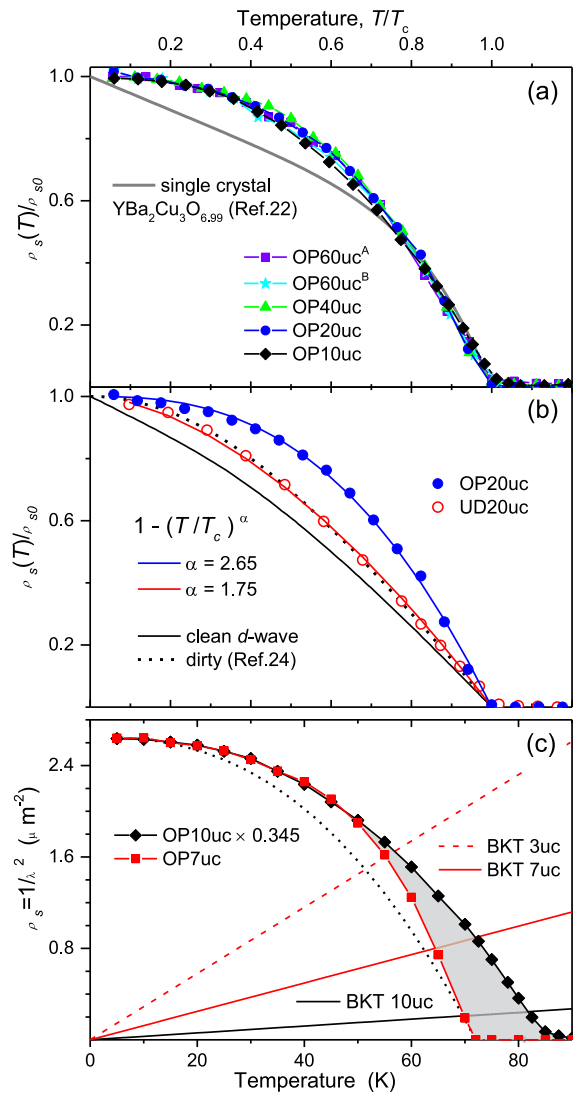}
                    \caption{(a) The temperature dependence of the normalized
superfluid density for five near-optimally doped DyBCO films with thicknesses
10 to 60 u.c. The solid gray line shows $\rho_s(T)/\rho_{s0}$ averaged over $a$ and $b$ crystallographic directions for pure YBa\ci{2}Cu\ci{3}O\ci{6.99} single crystals \cite{Bonn1998}. The values of $\rho_{s0}$ for all films are listed in Table I. (b) The normalized superfluid density for an  annealed underdoped 20\uc DyBCO film (red open circles). $\rho_s (T)/\rho_{s0}$ for this underdoped film agrees well with calculations for a dirty $d$-wave superconductor with a circular Fermi surface \cite{LeeHone2017} (black dotted line). (c) The superfluid density for near-optimally
doped 10 \uc (black diamonds) and 7 \uc (red squares) DyBCO films, with
$\rho_s (T)$ for the 10 \uc film scaled by 0.345. The linear dashed red, solid red, and solid black lines correspond to the expected BKT superfluid density for a 3 u.c., 7 \uc and 10 \uc thick SC layer, respectively, as described in the text. The black dotted curve in (c) represents the universal temperature dependence observed in (a).\label{SFD}}
                \end{figure}

To draw a comparison to the effects of underdoping, $\rho_s(T)/\rho_{s0}$ for a near-optimally doped 20 \uc film (OP20uc) is shown with that for annealed UD20uc in Fig. 2(b). In the case of the underdoped sample $\rho_s$ deviates from the universal trend observed in Fig. 2(a) and instead becomes more linear in temperature. To quantify this difference, we use a phenomenological fit of the measured curves of the form $1-(T/T_c)^\alpha$. For the reduced UD20uc film, we find a decrease in the scaling factor to $\alpha = 1.75$ from $\alpha = 2.65$ for OP20uc. The observed trend is consistent with previous observations for highly underdoped YBCO \cite{Zuev2005, Broun2007}, where deviations from the 3D-XY critical behavior were discussed as resulting from fluctuations near a quantum critical point approaching the (3+1)D-XY universality class \cite{Broun2007}. At the same time, the temperature dependence $\rho_s(T)/\rho_{s0}$ exhibits striking resemblance to calculations for a dirty $d$-wave superconductor with a circular Fermi surface of tetragonal symmetry \cite{LeeHone2017}. This agreement is illustrated by the comparison of the data to the black solid and dashed lines in Fig. 2(b), which represent calculations in the clean and dirty limit, respectively, with the unitarity-limit scattering $\Gamma_N = 0.1 T_{c0}$.
%, respectively, where the degree of scattering is characterized by the normal-state %scattering rate $\Gamma_N$ in units of the clean limit transition temperature $%T_{c0}$ for scatterers actingin the unitary limit.
The disorder causes a crossover from $T$-linear to $T^2$ behavior of $\rho_s(T)$ in the low temperature limit, whereas the particular choice of the Fermi surface determines the critical behavior as $T$ approaches $T_c$. The agreement between our data and the calculations are also consistent with a change of the Fermi surface topology from orthorhombic to tetragonal symmetry upon crossing over from near optimal doping to underdoped. Comparison of calculations to the entire temperature dependence of $\rho_s(T)/\rho_{s0}$ for the near-optimally doped films remains difficult, however, due to the incorporation of rectangular distortions of the Fermi surface. Further calculations based on the realistic Fermi surface of DyBCO are required to explain the temperature dependence of our universal behavior in Fig. 2(a).

We now turn to the central issue, the evolution of the superfluid density $\rho_s(T)$ when the film thickness is further reduced below 10 \uc For the near-optimally doped OP7uc shown by the red squares in Fig. 2(c), the steepness of $\rho_s(T)$ increases compared to the universal dependence, depicted by the black dotted line. Remarkably, when $\rho_s(T)$ for OP7uc is compared with the normalized superfluid density of OP10uc, it becomes evident that the two follow the same temperature dependence below 50 K. Above 50 K, however, $\rho_s(T)$ of OP7uc falls rapidly to zero, suppressing $T_c$ by $\approx$ 13 K with only minor broadening of the SC transition $\Delta
T_c$ from 1.4 K in OP10uc to 2 K in OP7uc, as defined by the FWHM of the peak in ImMI($T$) \cite{SM}. Such a reduction of $T_c$ is generally expected as a consequence of enhanced thermal fluctuations in two dimensions. In 2D superconductors, one expects a BKT transition with an abrupt (universal) jump in $\rho_s(T)$ at the temperature where this quantity intersects the line  $\rho_s(T_{\rm BKT}) = 8 \pi \mu_0k_B T / L \Phi_0^2$, where $L$ is the film thickness and $\Phi_0=h/2e$ is the flux quantum \cite{Kosterlitz2016,Minnhagen1987}. This line is shown in Fig. 2(c) for $L=$ 10 \uc (black solid line) and $L=$ 7 \uc (red solid line). Whereas $T_{BKT}$ is expected  very close to the measured $T_c$ for $L=$ 10 u.c., the BKT transition for $L=$ 7 \uc is predicted at a significantly lower temperature that is consistent with the temperature range in which $\rho_s(T)$ exhibits a steeper drop. However, instead of a genuine jump $\rho_s(T\rightarrow T_c)$ is rounded, implying the existence of $T_c$ inhomogeneity in the film. Our analysis of the charge carrier scattering shows that even our thinnest films remain in the moderate disorder regime, where the mean free path significantly exceeds the in-plane SC coherence length, $\ell/\xi_{ab}>1$ \cite{SM}. According to recent theoretical calculations \cite{Benfatto2008,Benfatto2013} the BKT jump is smeared by $T_c$ inhomogeneity around the average $T_{BKT}$, in very good agreement with our observations (cf. Fig. 4 in Ref.\cite{Benfatto2013}).

                \begin{table}
                        \begin{ruledtabular}
                        \begin{tabular}{ccccc}
                           
                            Film ID & L & $T_c$ & $\sigma_{dc}$&$\rho_{s0}$ \\
&(\uc/nm)&(K)&($m\Omega^{-1}cm^{-1}$)&($\mu m^{-2}$)\\

                            \hline
                            
                            OP60uc$^A$ &60/69.6 &  90    &   9.7       &
19.1 \\
                            OP60uc$^B$  &60/69.6 &  84    &   10.7       &
15.6 \\
                            OP40uc & 40/46.4 & 90 & 10.0 &16.0 \\
                            OP20uc & 20/23.3 & 82 & 16.4 &19.9 \\
                            OP10uc & 10/11.7 & 85 & 5.1 &7.67\\
                            OP7uc & 7/8.3 & 72 & 3.6 &2.65\\
                            UD20uc & 20/23.6 & 52 & 1.7 & 1.33 \\       
                \end{tabular}
                    \end{ruledtabular}
                    \caption{Nominal thicknesses $L$ and values of $T_c$,
zero-frequency limit of the real part of the optical conductivity at $T_c$,
 $\sigma_{dc}=\sigma_1(\omega \rightarrow
0,T_c)$, and zero-temperature superfluid density $\rho_{s0}$ obtained from
the data in Fig. 1.\label{AllFilms}}
                \end{table}
It remains plausible that our films are not uniformly superconducting throughout their entire thickness. To address this issue, in Fig. 3(a) we compare the critical temperature measured by MI with the film thickness and find that $T_c$ drops rapidly to zero near 4 \uc \cite{Putzky2020}. Due to the large variation in interfacial and impurity scattering exhibited by the various films, we cannot directly draw a comparison between $\rho_{s0}$ and thickness $L$. Rather, we normalize $\rho_{s0}$ by the normal state conductivity $\sigma_{dc}(T_c)$ and $T_c$ \cite{Homes2004}. Curiously, the quantity $\rho_{s0}L/\sigma_{dc} T_c$, proportional to the effective thickness of the SC layer, trends linearly to zero also at 4 u.c., suggesting that at this thickness the fraction of SC material in the film drops to zero. This implies that a non-SC interfacial layer of thickness $\sim 4$ \uc persists in all DyBCO films. At a thickness of 7 \uc only a thin 3 \uc SC layer remains which approaches the threshold for 2D-SC fluctuations to emerge, as schematically presented in Fig. 3(b). The red dashed line in Fig. 2(c) represents the BKT line with the effective SC layer thickness reduced to 3 u.c. Assuming that the superfluid density should also be normalized by the effective thickness of the SC layer, this dashed line serves as a lower bound for the BKT transition temperature in OP7uc.

  \begin{figure}
                    \includegraphics[width=76mm]{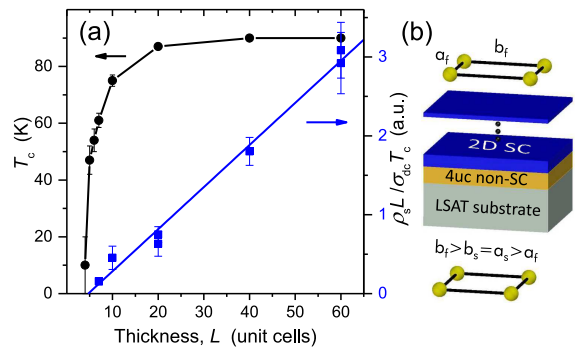}
                    \caption{(a) The thickness dependence of of $T_c$ of the \mbox{DyBCO} films (solid circles, left axis) as determined by MI measurements \cite{Putzky2020}. Samples thinner than 5 u.c. were unstable and displayed a broad ImMI($T$) peak close to $T=0$. Also plotted as a function of thickness
is the quantity $\rho_{s0}L/\sigma_{dc}T_c$, which represents a measure of the
effective SC layer thickness (blue squares, right axis). (b) Schematic representation of the DyBCO films wherein a 4 \uc non-SC layer sits at the substrate-film interface to adapt the orthorhombic structure of the
DyBCO film ($b_f>a_f$) to the square lattice of the underlying LSAT substrate ($a_s=b_s$).\label{Tc}}
                \end{figure}
                
The non-SC 4 \uc interfacial layer plays the role of the semiconducting (Pr,Y)BCO buffer layer used in previous studies of ultrathin SC YBCO films \cite{Hetel2007,Matsuda1993, Repaci1996,Gasparov2005,Varela1999,Chan1993} to adapt the orthorhombic structure of the DyBCO film to the square lattice of the underlying substrate. 
Indeed, $T_c(L-4$ u.c.$)$ in our films reproduces quite well $T_c(L)$ observed in YBCO/(Pr,Y)BCO multilayers \cite{Varela1999,Chan1993}.
The relatively homogeneous and large $T_c$ even in films as thin as 7 u.c. imply that the interfacial layer is sharp and well-defined.
The absence of 4 eV peak in Fig. 1(c) in all our samples independent of thickness suggests that the films are uniformly oxygenated  and  an epitaxially induced oxygen vacancy distribution is not primarily responsible for the origin of the non-SC layer. Epitaxially stabilized competing order, such as charge density wave correlations, which universally exist in cuprate HTSCs \cite{Keimer2015}, may instead be present and suppress SC. Recent resonant X-ray scattering measurements on underdoped YBCO films grown epitaxially on STO indicate that epitaxial strain can stabilize 3D charge order (CO), with the Cu sites in the CuO chain layers participating in the CO state \cite{Bluschke2018}. There, the apparent lack of competition between 3D-CO and superconductivity was interpreted as possible evidence of mesoscopic phase separation between regions hosting the CO and SC states. It is therefore possible that the 4 \uc interfacial layer represents an intrinsic CO but non-SC region, which may remain metallic and provides an avenue to understanding recent reports of an anomalous bosonic metallic state in 10 u.c. SC YBCO films \cite{Yang2019}.

In summary, we have carefully examined the temperature dependence of the superfluid density for a series of ultrathin DyBCO films. We find that films 10 u.c. and thicker near optimal doping display a universal temperature dependence of $\rho_s(T)/\rho_{s0}$. This universality allows us to draw a clear distinction with the behavior of $\rho_s(T)$ in a thinner 7 u.c. sample, where the steepness of $\rho_s(T)$ as $T$ approaches $T_c$ markedly increases. We assign the suppression of $T_c$ in this film to the emergence of 2D-SC fluctuations. Further studies of the magnetic field dependence of $\rho_s(T)$ are important to elucidate this scenario. Our analysis suggests that a sharply defined 4 u.c. thick interfacial layer remains non-SC, leaving a quasi-2D-SC layer on top. We propose that this interfacial layer hosts an epitaxially stabilized non-SC CO state. Our results provide a promising platform to study 2D-SC and its interplay with CO in cuprate HTSCs.

\section{Acknowledgments}
        We thank S.~A. Kivelson, D.~V. Efremov, N.~R. Lee-Hone, J.~S. Dodge, and G. Kim for fruitful discussions. We gratefully acknowledge Y.-L. Mathis for support at the IR1 beamline of the Karlsruhe Research Accelerator (KARA). 
%This work was partly supported by ....

\nocite{Chiao2000,Tomimoto1999,Albenque2003}

%\bibliography{2DDyBCO24Sep2020}

\begin{thebibliography}{35}%
\makeatletter
\providecommand \@ifxundefined [1]{%
 \@ifx{#1\undefined}
}%
\providecommand \@ifnum [1]{%
 \ifnum #1\expandafter \@firstoftwo
 \else \expandafter \@secondoftwo
 \fi
}%
\providecommand \@ifx [1]{%
 \ifx #1\expandafter \@firstoftwo
 \else \expandafter \@secondoftwo
 \fi
}%
\providecommand \natexlab [1]{#1}%
\providecommand \enquote  [1]{``#1''}%
\providecommand \bibnamefont  [1]{#1}%
\providecommand \bibfnamefont [1]{#1}%
\providecommand \citenamefont [1]{#1}%
\providecommand \href@noop [0]{\@secondoftwo}%
\providecommand \href [0]{\begingroup \@sanitize@url \@href}%
\providecommand \@href[1]{\@@startlink{#1}\@@href}%
\providecommand \@@href[1]{\endgroup#1\@@endlink}%
\providecommand \@sanitize@url [0]{\catcode `\\12\catcode `\$12\catcode
  `\&12\catcode `\#12\catcode `\^12\catcode `\_12\catcode `\%12\relax}%
\providecommand \@@startlink[1]{}%
\providecommand \@@endlink[0]{}%
\providecommand \url  [0]{\begingroup\@sanitize@url \@url }%
\providecommand \@url [1]{\endgroup\@href {#1}{\urlprefix }}%
\providecommand \urlprefix  [0]{URL }%
\providecommand \Eprint [0]{\href }%
\providecommand \doibase [0]{http://dx.doi.org/}%
\providecommand \selectlanguage [0]{\@gobble}%
\providecommand \bibinfo  [0]{\@secondoftwo}%
\providecommand \bibfield  [0]{\@secondoftwo}%
\providecommand \translation [1]{[#1]}%
\providecommand \BibitemOpen [0]{}%
\providecommand \bibitemStop [0]{}%
\providecommand \bibitemNoStop [0]{.\EOS\space}%
\providecommand \EOS [0]{\spacefactor3000\relax}%
\providecommand \BibitemShut  [1]{\csname bibitem#1\endcsname}%
\let\auto@bib@innerbib\@empty
%</preamble>
\bibitem [{\citenamefont {Kamal}\ \emph {et~al.}(1994)\citenamefont {Kamal},
  \citenamefont {Bonn}, \citenamefont {Goldenfeld}, \citenamefont {Hirschfeld},
  \citenamefont {Liang},\ and\ \citenamefont {Hardy}}]{BonnPRL1994}%
  \BibitemOpen
  \bibfield  {author} {\bibinfo {author} {\bibfnamefont {S.}~\bibnamefont
  {Kamal}}, \bibinfo {author} {\bibfnamefont {D.~A.}\ \bibnamefont {Bonn}},
  \bibinfo {author} {\bibfnamefont {N.}~\bibnamefont {Goldenfeld}}, \bibinfo
  {author} {\bibfnamefont {P.~J.}\ \bibnamefont {Hirschfeld}}, \bibinfo
  {author} {\bibfnamefont {R.}~\bibnamefont {Liang}}, \ and\ \bibinfo {author}
  {\bibfnamefont {W.~N.}\ \bibnamefont {Hardy}},\ }\href {\doibase
  10.1103/PhysRevLett.73.1845} {\bibfield  {journal} {\bibinfo  {journal}
  {Phys. Rev. Lett.}\ }\textbf {\bibinfo {volume} {73}},\ \bibinfo {pages}
  {1845} (\bibinfo {year} {1994})}\BibitemShut {NoStop}%
\bibitem [{\citenamefont {Meingast}\ \emph {et~al.}(2001)\citenamefont
  {Meingast}, \citenamefont {Pasler}, \citenamefont {Nagel}, \citenamefont
  {Rykov}, \citenamefont {Tajima},\ and\ \citenamefont
  {Olsson}}]{Meingast2001}%
  \BibitemOpen
  \bibfield  {author} {\bibinfo {author} {\bibfnamefont {C.}~\bibnamefont
  {Meingast}}, \bibinfo {author} {\bibfnamefont {V.}~\bibnamefont {Pasler}},
  \bibinfo {author} {\bibfnamefont {P.}~\bibnamefont {Nagel}}, \bibinfo
  {author} {\bibfnamefont {A.}~\bibnamefont {Rykov}}, \bibinfo {author}
  {\bibfnamefont {S.}~\bibnamefont {Tajima}}, \ and\ \bibinfo {author}
  {\bibfnamefont {P.}~\bibnamefont {Olsson}},\ }\href {\doibase
  10.1103/PhysRevLett.86.1606} {\bibfield  {journal} {\bibinfo  {journal}
  {Phys. Rev. Lett.}\ }\textbf {\bibinfo {volume} {86}},\ \bibinfo {pages}
  {1606} (\bibinfo {year} {2001})}\BibitemShut {NoStop}%
\bibitem [{\citenamefont {Grbi\'{c}}\ \emph {et~al.}(2011)\citenamefont
  {Grbi\'{c}}, \citenamefont {Po\v{z}ek}, \citenamefont {Paar}, \citenamefont
  {Hinkov}, \citenamefont {Raichle}, \citenamefont {Haug}, \citenamefont
  {Keimer}, \citenamefont {Bari\v{s}i\'{c}},\ and\ \citenamefont
  {Dul\v{c}i\'{c}}}]{Grbic2011}%
  \BibitemOpen
  \bibfield  {author} {\bibinfo {author} {\bibfnamefont {M.~S.}\ \bibnamefont
  {Grbi\'{c}}}, \bibinfo {author} {\bibfnamefont {M.}~\bibnamefont
  {Po\v{z}ek}}, \bibinfo {author} {\bibfnamefont {D.}~\bibnamefont {Paar}},
  \bibinfo {author} {\bibfnamefont {V.}~\bibnamefont {Hinkov}}, \bibinfo
  {author} {\bibfnamefont {M.}~\bibnamefont {Raichle}}, \bibinfo {author}
  {\bibfnamefont {D.}~\bibnamefont {Haug}}, \bibinfo {author} {\bibfnamefont
  {B.}~\bibnamefont {Keimer}}, \bibinfo {author} {\bibfnamefont
  {N.}~\bibnamefont {Bari\v{s}i\'{c}}}, \ and\ \bibinfo {author} {\bibfnamefont
  {A.}~\bibnamefont {Dul\v{c}i\'{c}}},\ }\href {\doibase
  10.1103/PhysRevB.83.144508} {\bibfield  {journal} {\bibinfo  {journal} {Phys.
  Rev. B}\ }\textbf {\bibinfo {volume} {83}},\ \bibinfo {pages} {144508}
  (\bibinfo {year} {2011})}\BibitemShut {NoStop}%
\bibitem [{\citenamefont {Nizhankovskiy}\ and\ \citenamefont
  {Rogacki}(2019)}]{Nizhankovskiy2019}%
  \BibitemOpen
  \bibfield  {author} {\bibinfo {author} {\bibfnamefont {V.~I.}\ \bibnamefont
  {Nizhankovskiy}}\ and\ \bibinfo {author} {\bibfnamefont {K.}~\bibnamefont
  {Rogacki}},\ }\href {\doibase 10.1103/PhysRevB.100.104510} {\bibfield
  {journal} {\bibinfo  {journal} {Phys. Rev. B}\ }\textbf {\bibinfo {volume}
  {100}},\ \bibinfo {pages} {104510} (\bibinfo {year} {2019})}\BibitemShut
  {NoStop}%
\bibitem [{\citenamefont {Kosterlitz}(2016)}]{Kosterlitz2016}%
  \BibitemOpen
  \bibfield  {author} {\bibinfo {author} {\bibfnamefont {J.~M.}\ \bibnamefont
  {Kosterlitz}},\ }\href {\doibase 10.1088/0034-4885/79/2/026001} {\bibfield
  {journal} {\bibinfo  {journal} {Reports on Progress in Physics}\ }\textbf
  {\bibinfo {volume} {79}},\ \bibinfo {pages} {026001} (\bibinfo {year}
  {2016})}\BibitemShut {NoStop}%
\bibitem [{\citenamefont {Minnhagen}(1987)}]{Minnhagen1987}%
  \BibitemOpen
  \bibfield  {author} {\bibinfo {author} {\bibfnamefont {P.}~\bibnamefont
  {Minnhagen}},\ }\href {\doibase 10.1103/RevModPhys.59.1001} {\bibfield
  {journal} {\bibinfo  {journal} {Rev. Mod. Phys.}\ }\textbf {\bibinfo {volume}
  {59}},\ \bibinfo {pages} {1001} (\bibinfo {year} {1987})}\BibitemShut
  {NoStop}%
\bibitem [{\citenamefont {Cyr-Choini\`ere}\ \emph {et~al.}(2018)\citenamefont
  {Cyr-Choini\`ere}, \citenamefont {Daou}, \citenamefont {Lalibert\'e},
  \citenamefont {Collignon}, \citenamefont {Badoux}, \citenamefont {LeBoeuf},
  \citenamefont {Chang}, \citenamefont {Ramshaw}, \citenamefont {Bonn},
  \citenamefont {Hardy}, \citenamefont {Liang}, \citenamefont {Yan},
  \citenamefont {Cheng}, \citenamefont {Zhou}, \citenamefont {Goodenough},
  \citenamefont {Pyon}, \citenamefont {Takayama}, \citenamefont {Takagi},
  \citenamefont {Doiron-Leyraud},\ and\ \citenamefont {Taillefer}}]{Cyr2018}%
  \BibitemOpen
  \bibfield  {author} {\bibinfo {author} {\bibfnamefont {O.}~\bibnamefont
  {Cyr-Choini\`ere}}, \bibinfo {author} {\bibfnamefont {R.}~\bibnamefont
  {Daou}}, \bibinfo {author} {\bibfnamefont {F.}~\bibnamefont {Lalibert\'e}},
  \bibinfo {author} {\bibfnamefont {C.}~\bibnamefont {Collignon}}, \bibinfo
  {author} {\bibfnamefont {S.}~\bibnamefont {Badoux}}, \bibinfo {author}
  {\bibfnamefont {D.}~\bibnamefont {LeBoeuf}}, \bibinfo {author} {\bibfnamefont
  {J.}~\bibnamefont {Chang}}, \bibinfo {author} {\bibfnamefont {B.~J.}\
  \bibnamefont {Ramshaw}}, \bibinfo {author} {\bibfnamefont {D.~A.}\
  \bibnamefont {Bonn}}, \bibinfo {author} {\bibfnamefont {W.~N.}\ \bibnamefont
  {Hardy}}, \bibinfo {author} {\bibfnamefont {R.}~\bibnamefont {Liang}},
  \bibinfo {author} {\bibfnamefont {J.-Q.}\ \bibnamefont {Yan}}, \bibinfo
  {author} {\bibfnamefont {J.-G.}\ \bibnamefont {Cheng}}, \bibinfo {author}
  {\bibfnamefont {J.-S.}\ \bibnamefont {Zhou}}, \bibinfo {author}
  {\bibfnamefont {J.~B.}\ \bibnamefont {Goodenough}}, \bibinfo {author}
  {\bibfnamefont {S.}~\bibnamefont {Pyon}}, \bibinfo {author} {\bibfnamefont
  {T.}~\bibnamefont {Takayama}}, \bibinfo {author} {\bibfnamefont
  {H.}~\bibnamefont {Takagi}}, \bibinfo {author} {\bibfnamefont
  {N.}~\bibnamefont {Doiron-Leyraud}}, \ and\ \bibinfo {author} {\bibfnamefont
  {L.}~\bibnamefont {Taillefer}},\ }\href {\doibase 10.1103/PhysRevB.97.064502}
  {\bibfield  {journal} {\bibinfo  {journal} {Phys. Rev. B}\ }\textbf {\bibinfo
  {volume} {97}},\ \bibinfo {pages} {064502} (\bibinfo {year}
  {2018})}\BibitemShut {NoStop}%
\bibitem [{\citenamefont {Dubroka}\ \emph {et~al.}(2011)\citenamefont
  {Dubroka}, \citenamefont {R\"ossle}, \citenamefont {Kim}, \citenamefont
  {Malik}, \citenamefont {Munzar}, \citenamefont {Basov}, \citenamefont
  {Schafgans}, \citenamefont {Moon}, \citenamefont {Lin}, \citenamefont {Haug},
  \citenamefont {Hinkov}, \citenamefont {Keimer}, \citenamefont {Wolf},
  \citenamefont {Storey}, \citenamefont {Tallon},\ and\ \citenamefont
  {Bernhard}}]{Dubroka2011}%
  \BibitemOpen
  \bibfield  {author} {\bibinfo {author} {\bibfnamefont {A.}~\bibnamefont
  {Dubroka}}, \bibinfo {author} {\bibfnamefont {M.}~\bibnamefont {R\"ossle}},
  \bibinfo {author} {\bibfnamefont {K.~W.}\ \bibnamefont {Kim}}, \bibinfo
  {author} {\bibfnamefont {V.~K.}\ \bibnamefont {Malik}}, \bibinfo {author}
  {\bibfnamefont {D.}~\bibnamefont {Munzar}}, \bibinfo {author} {\bibfnamefont
  {D.~N.}\ \bibnamefont {Basov}}, \bibinfo {author} {\bibfnamefont {A.~A.}\
  \bibnamefont {Schafgans}}, \bibinfo {author} {\bibfnamefont {S.~J.}\
  \bibnamefont {Moon}}, \bibinfo {author} {\bibfnamefont {C.~T.}\ \bibnamefont
  {Lin}}, \bibinfo {author} {\bibfnamefont {D.}~\bibnamefont {Haug}}, \bibinfo
  {author} {\bibfnamefont {V.}~\bibnamefont {Hinkov}}, \bibinfo {author}
  {\bibfnamefont {B.}~\bibnamefont {Keimer}}, \bibinfo {author} {\bibfnamefont
  {T.}~\bibnamefont {Wolf}}, \bibinfo {author} {\bibfnamefont {J.~G.}\
  \bibnamefont {Storey}}, \bibinfo {author} {\bibfnamefont {J.~L.}\
  \bibnamefont {Tallon}}, \ and\ \bibinfo {author} {\bibfnamefont
  {C.}~\bibnamefont {Bernhard}},\ }\href {\doibase
  10.1103/PhysRevLett.106.047006} {\bibfield  {journal} {\bibinfo  {journal}
  {Phys. Rev. Lett.}\ }\textbf {\bibinfo {volume} {106}},\ \bibinfo {pages}
  {047006} (\bibinfo {year} {2011})}\BibitemShut {NoStop}%
\bibitem [{\citenamefont {Broun}\ \emph {et~al.}(2007)\citenamefont {Broun},
  \citenamefont {Huttema}, \citenamefont {Turner}, \citenamefont {\"Ozcan},
  \citenamefont {Morgan}, \citenamefont {Liang}, \citenamefont {Hardy},\ and\
  \citenamefont {Bonn}}]{Broun2007}%
  \BibitemOpen
  \bibfield  {author} {\bibinfo {author} {\bibfnamefont {D.~M.}\ \bibnamefont
  {Broun}}, \bibinfo {author} {\bibfnamefont {W.~A.}\ \bibnamefont {Huttema}},
  \bibinfo {author} {\bibfnamefont {P.~J.}\ \bibnamefont {Turner}}, \bibinfo
  {author} {\bibfnamefont {S.}~\bibnamefont {\"Ozcan}}, \bibinfo {author}
  {\bibfnamefont {B.}~\bibnamefont {Morgan}}, \bibinfo {author} {\bibfnamefont
  {R.}~\bibnamefont {Liang}}, \bibinfo {author} {\bibfnamefont {W.~N.}\
  \bibnamefont {Hardy}}, \ and\ \bibinfo {author} {\bibfnamefont {D.~A.}\
  \bibnamefont {Bonn}},\ }\href {\doibase 10.1103/PhysRevLett.99.237003}
  {\bibfield  {journal} {\bibinfo  {journal} {Phys. Rev. Lett.}\ }\textbf
  {\bibinfo {volume} {99}},\ \bibinfo {pages} {237003} (\bibinfo {year}
  {2007})}\BibitemShut {NoStop}%
\bibitem [{\citenamefont {Blatter}\ \emph {et~al.}(1994)\citenamefont
  {Blatter}, \citenamefont {Feigel'man}, \citenamefont {Geshkenbein},
  \citenamefont {Larkin},\ and\ \citenamefont {Vinokur}}]{Blatter1994}%
  \BibitemOpen
  \bibfield  {author} {\bibinfo {author} {\bibfnamefont {G.}~\bibnamefont
  {Blatter}}, \bibinfo {author} {\bibfnamefont {M.~V.}\ \bibnamefont
  {Feigel'man}}, \bibinfo {author} {\bibfnamefont {V.~B.}\ \bibnamefont
  {Geshkenbein}}, \bibinfo {author} {\bibfnamefont {A.~I.}\ \bibnamefont
  {Larkin}}, \ and\ \bibinfo {author} {\bibfnamefont {V.~M.}\ \bibnamefont
  {Vinokur}},\ }\href {\doibase 10.1103/RevModPhys.66.1125} {\bibfield
  {journal} {\bibinfo  {journal} {Rev. Mod. Phys.}\ }\textbf {\bibinfo {volume}
  {66}},\ \bibinfo {pages} {1125} (\bibinfo {year} {1994})}\BibitemShut
  {NoStop}%
\bibitem [{\citenamefont {Keimer}\ \emph {et~al.}(2015)\citenamefont {Keimer},
  \citenamefont {Kivelson}, \citenamefont {Norman}, \citenamefont {Uchida},\
  and\ \citenamefont {Zaanen}}]{Keimer2015}%
  \BibitemOpen
  \bibfield  {author} {\bibinfo {author} {\bibfnamefont {B.}~\bibnamefont
  {Keimer}}, \bibinfo {author} {\bibfnamefont {S.~A.}\ \bibnamefont
  {Kivelson}}, \bibinfo {author} {\bibfnamefont {M.~R.}\ \bibnamefont
  {Norman}}, \bibinfo {author} {\bibfnamefont {S.}~\bibnamefont {Uchida}}, \
  and\ \bibinfo {author} {\bibfnamefont {J.}~\bibnamefont {Zaanen}},\ }\href
  {\doibase 10.1038/nature14165} {\bibfield  {journal} {\bibinfo  {journal}
  {Nature}\ }\textbf {\bibinfo {volume} {518}},\ \bibinfo {pages} {179}
  (\bibinfo {year} {2015})}\BibitemShut {NoStop}%
\bibitem [{\citenamefont {Matsuda}\ \emph {et~al.}(1993)\citenamefont
  {Matsuda}, \citenamefont {Komiyama}, \citenamefont {Onogi}, \citenamefont
  {Terashima}, \citenamefont {Shimura},\ and\ \citenamefont
  {Bando}}]{Matsuda1993}%
  \BibitemOpen
  \bibfield  {author} {\bibinfo {author} {\bibfnamefont {Y.}~\bibnamefont
  {Matsuda}}, \bibinfo {author} {\bibfnamefont {S.}~\bibnamefont {Komiyama}},
  \bibinfo {author} {\bibfnamefont {T.}~\bibnamefont {Onogi}}, \bibinfo
  {author} {\bibfnamefont {T.}~\bibnamefont {Terashima}}, \bibinfo {author}
  {\bibfnamefont {K.}~\bibnamefont {Shimura}}, \ and\ \bibinfo {author}
  {\bibfnamefont {Y.}~\bibnamefont {Bando}},\ }\href {\doibase
  10.1103/PhysRevB.48.10498} {\bibfield  {journal} {\bibinfo  {journal} {Phys.
  Rev. B}\ }\textbf {\bibinfo {volume} {48}},\ \bibinfo {pages} {10498}
  (\bibinfo {year} {1993})}\BibitemShut {NoStop}%
\bibitem [{\citenamefont {Repaci}\ \emph {et~al.}(1996)\citenamefont {Repaci},
  \citenamefont {Kwon}, \citenamefont {Li}, \citenamefont {Jiang},
  \citenamefont {Venkatessan}, \citenamefont {Glover}, \citenamefont {Lobb},\
  and\ \citenamefont {Newrock}}]{Repaci1996}%
  \BibitemOpen
  \bibfield  {author} {\bibinfo {author} {\bibfnamefont {J.~M.}\ \bibnamefont
  {Repaci}}, \bibinfo {author} {\bibfnamefont {C.}~\bibnamefont {Kwon}},
  \bibinfo {author} {\bibfnamefont {Q.}~\bibnamefont {Li}}, \bibinfo {author}
  {\bibfnamefont {X.}~\bibnamefont {Jiang}}, \bibinfo {author} {\bibfnamefont
  {T.}~\bibnamefont {Venkatessan}}, \bibinfo {author} {\bibfnamefont {R.~E.}\
  \bibnamefont {Glover}}, \bibinfo {author} {\bibfnamefont {C.~J.}\
  \bibnamefont {Lobb}}, \ and\ \bibinfo {author} {\bibfnamefont {R.~S.}\
  \bibnamefont {Newrock}},\ }\href {\doibase 10.1103/PhysRevB.54.R9674}
  {\bibfield  {journal} {\bibinfo  {journal} {Phys. Rev. B}\ }\textbf {\bibinfo
  {volume} {54}},\ \bibinfo {pages} {R9674} (\bibinfo {year}
  {1996})}\BibitemShut {NoStop}%
\bibitem [{\citenamefont {Gasparov}\ \emph {et~al.}(2005)\citenamefont
  {Gasparov}, \citenamefont {Tsydynzhapov}, \citenamefont {Batov},\ and\
  \citenamefont {Li}}]{Gasparov2005}%
  \BibitemOpen
  \bibfield  {author} {\bibinfo {author} {\bibfnamefont {V.}~\bibnamefont
  {Gasparov}}, \bibinfo {author} {\bibfnamefont {G.}~\bibnamefont
  {Tsydynzhapov}}, \bibinfo {author} {\bibfnamefont {I.}~\bibnamefont {Batov}},
  \ and\ \bibinfo {author} {\bibfnamefont {Q.}~\bibnamefont {Li}},\ }\href@noop
  {} {\bibfield  {journal} {\bibinfo  {journal} {Journal of Low Temperature
  Physics}\ }\textbf {\bibinfo {volume} {139}},\ \bibinfo {pages} {49}
  (\bibinfo {year} {2005})}\BibitemShut {NoStop}%
\bibitem [{\citenamefont {Hetel}\ \emph {et~al.}(2007)\citenamefont {Hetel},
  \citenamefont {Lemberger},\ and\ \citenamefont {Randeria}}]{Hetel2007}%
  \BibitemOpen
  \bibfield  {author} {\bibinfo {author} {\bibfnamefont {I.}~\bibnamefont
  {Hetel}}, \bibinfo {author} {\bibfnamefont {T.~R.}\ \bibnamefont
  {Lemberger}}, \ and\ \bibinfo {author} {\bibfnamefont {M.}~\bibnamefont
  {Randeria}},\ }\href {\doibase 10.1038/nphys707} {\bibfield  {journal}
  {\bibinfo  {journal} {Nature Physics}\ }\textbf {\bibinfo {volume} {3}},\
  \bibinfo {pages} {700} (\bibinfo {year} {2007})}\BibitemShut {NoStop}%
\bibitem [{\citenamefont {Putzky}\ \emph {et~al.}(2020)\citenamefont {Putzky},
  \citenamefont {Radhakrishnan}, \citenamefont {Wang}, \citenamefont {Wochner},
  \citenamefont {Christiani}, \citenamefont {Minola}, \citenamefont {van Aken},
  \citenamefont {Logvenov}, \citenamefont {Benckiser},\ and\ \citenamefont
  {Keimer}}]{Putzky2020}%
  \BibitemOpen
  \bibfield  {author} {\bibinfo {author} {\bibfnamefont {D.}~\bibnamefont
  {Putzky}}, \bibinfo {author} {\bibfnamefont {P.}~\bibnamefont
  {Radhakrishnan}}, \bibinfo {author} {\bibfnamefont {Y.}~\bibnamefont {Wang}},
  \bibinfo {author} {\bibfnamefont {P.}~\bibnamefont {Wochner}}, \bibinfo
  {author} {\bibfnamefont {G.}~\bibnamefont {Christiani}}, \bibinfo {author}
  {\bibfnamefont {M.}~\bibnamefont {Minola}}, \bibinfo {author} {\bibfnamefont
  {P.~A.}\ \bibnamefont {van Aken}}, \bibinfo {author} {\bibfnamefont
  {G.}~\bibnamefont {Logvenov}}, \bibinfo {author} {\bibfnamefont
  {E.}~\bibnamefont {Benckiser}}, \ and\ \bibinfo {author} {\bibfnamefont
  {B.}~\bibnamefont {Keimer}},\ }\href {\doibase 10.1063/5.0019673} {\bibfield
  {journal} {\bibinfo  {journal} {Applied Physics Letters}\ }\textbf {\bibinfo
  {volume} {117}},\ \bibinfo {pages} {072601} (\bibinfo {year}
  {2020})}\BibitemShut {NoStop}%
\bibitem{SM} See Supplemental Material at DOI: 10.1103/PhysRevLett.125.237001 for more information about experimental and data analysis.
\bibitem [{\citenamefont {Kircher}\ \emph {et~al.}(1989)\citenamefont
  {Kircher}, \citenamefont {Alouani}, \citenamefont {Garriga}, \citenamefont
  {Murugaraj}, \citenamefont {Maier}, \citenamefont {Thomsen}, \citenamefont
  {Cardona}, \citenamefont {Andersen},\ and\ \citenamefont
  {Jepsen}}]{Kircher1989}%
  \BibitemOpen
  \bibfield  {author} {\bibinfo {author} {\bibfnamefont {J.}~\bibnamefont
  {Kircher}}, \bibinfo {author} {\bibfnamefont {M.}~\bibnamefont {Alouani}},
  \bibinfo {author} {\bibfnamefont {M.}~\bibnamefont {Garriga}}, \bibinfo
  {author} {\bibfnamefont {P.}~\bibnamefont {Murugaraj}}, \bibinfo {author}
  {\bibfnamefont {J.}~\bibnamefont {Maier}}, \bibinfo {author} {\bibfnamefont
  {C.}~\bibnamefont {Thomsen}}, \bibinfo {author} {\bibfnamefont
  {M.}~\bibnamefont {Cardona}}, \bibinfo {author} {\bibfnamefont {O.~K.}\
  \bibnamefont {Andersen}}, \ and\ \bibinfo {author} {\bibfnamefont
  {O.}~\bibnamefont {Jepsen}},\ }\href {\doibase 10.1103/PhysRevB.40.7368}
  {\bibfield  {journal} {\bibinfo  {journal} {Phys. Rev. B}\ }\textbf {\bibinfo
  {volume} {40}},\ \bibinfo {pages} {7368} (\bibinfo {year}
  {1989})}\BibitemShut {NoStop}%
\bibitem [{\citenamefont {Cooper}\ \emph {et~al.}(1992)\citenamefont {Cooper},
  \citenamefont {Kotz}, \citenamefont {Karlow}, \citenamefont {Klein},
  \citenamefont {Lee}, \citenamefont {Giapintzakis},\ and\ \citenamefont
  {Ginsberg}}]{Cooper1992}%
  \BibitemOpen
  \bibfield  {author} {\bibinfo {author} {\bibfnamefont {S.~L.}\ \bibnamefont
  {Cooper}}, \bibinfo {author} {\bibfnamefont {A.~L.}\ \bibnamefont {Kotz}},
  \bibinfo {author} {\bibfnamefont {M.~A.}\ \bibnamefont {Karlow}}, \bibinfo
  {author} {\bibfnamefont {M.~V.}\ \bibnamefont {Klein}}, \bibinfo {author}
  {\bibfnamefont {W.~C.}\ \bibnamefont {Lee}}, \bibinfo {author} {\bibfnamefont
  {J.}~\bibnamefont {Giapintzakis}}, \ and\ \bibinfo {author} {\bibfnamefont
  {D.~M.}\ \bibnamefont {Ginsberg}},\ }\href {\doibase
  10.1103/PhysRevB.45.2549} {\bibfield  {journal} {\bibinfo  {journal} {Phys.
  Rev. B}\ }\textbf {\bibinfo {volume} {45}},\ \bibinfo {pages} {2549}
  (\bibinfo {year} {1992})}\BibitemShut {NoStop}%
\bibitem [{\citenamefont {Hosseini}\ \emph {et~al.}(1999)\citenamefont
  {Hosseini}, \citenamefont {Harris}, \citenamefont {Kamal}, \citenamefont
  {Dosanjh}, \citenamefont {Preston}, \citenamefont {Liang}, \citenamefont
  {Hardy},\ and\ \citenamefont {Bonn}}]{Bonn1999}%
  \BibitemOpen
  \bibfield  {author} {\bibinfo {author} {\bibfnamefont {A.}~\bibnamefont
  {Hosseini}}, \bibinfo {author} {\bibfnamefont {R.}~\bibnamefont {Harris}},
  \bibinfo {author} {\bibfnamefont {S.}~\bibnamefont {Kamal}}, \bibinfo
  {author} {\bibfnamefont {P.}~\bibnamefont {Dosanjh}}, \bibinfo {author}
  {\bibfnamefont {J.}~\bibnamefont {Preston}}, \bibinfo {author} {\bibfnamefont
  {R.}~\bibnamefont {Liang}}, \bibinfo {author} {\bibfnamefont {W.~N.}\
  \bibnamefont {Hardy}}, \ and\ \bibinfo {author} {\bibfnamefont {D.~A.}\
  \bibnamefont {Bonn}},\ }\href {\doibase 10.1103/PhysRevB.60.1349} {\bibfield
  {journal} {\bibinfo  {journal} {Phys. Rev. B}\ }\textbf {\bibinfo {volume}
  {60}},\ \bibinfo {pages} {1349} (\bibinfo {year} {1999})}\BibitemShut
  {NoStop}%
\bibitem [{\citenamefont {Fink}(2000)}]{Fink2000}%
  \BibitemOpen
  \bibfield  {author} {\bibinfo {author} {\bibfnamefont {H.~J.}\ \bibnamefont
  {Fink}},\ }\href {\doibase 10.1103/PhysRevB.61.6346} {\bibfield  {journal}
  {\bibinfo  {journal} {Phys. Rev. B}\ }\textbf {\bibinfo {volume} {61}},\
  \bibinfo {pages} {6346} (\bibinfo {year} {2000})}\BibitemShut {NoStop}%
\bibitem [{\citenamefont {Kamal}\ \emph {et~al.}(1998)\citenamefont {Kamal},
  \citenamefont {Liang}, \citenamefont {Hosseini}, \citenamefont {Bonn},\ and\
  \citenamefont {Hardy}}]{Bonn1998}%
  \BibitemOpen
  \bibfield  {author} {\bibinfo {author} {\bibfnamefont {S.}~\bibnamefont
  {Kamal}}, \bibinfo {author} {\bibfnamefont {R.}~\bibnamefont {Liang}},
  \bibinfo {author} {\bibfnamefont {A.}~\bibnamefont {Hosseini}}, \bibinfo
  {author} {\bibfnamefont {D.~A.}\ \bibnamefont {Bonn}}, \ and\ \bibinfo
  {author} {\bibfnamefont {W.~N.}\ \bibnamefont {Hardy}},\ }\href {\doibase
  10.1103/PhysRevB.58.R8933} {\bibfield  {journal} {\bibinfo  {journal} {Phys.
  Rev. B}\ }\textbf {\bibinfo {volume} {58}},\ \bibinfo {pages} {R8933}
  (\bibinfo {year} {1998})}\BibitemShut {NoStop}%
\bibitem [{\citenamefont {Hardy}\ \emph {et~al.}(1993)\citenamefont {Hardy},
  \citenamefont {Bonn}, \citenamefont {Morgan}, \citenamefont {Liang},\ and\
  \citenamefont {Zhang}}]{Hardy1993}%
  \BibitemOpen
  \bibfield  {author} {\bibinfo {author} {\bibfnamefont {W.~N.}\ \bibnamefont
  {Hardy}}, \bibinfo {author} {\bibfnamefont {D.~A.}\ \bibnamefont {Bonn}},
  \bibinfo {author} {\bibfnamefont {D.~C.}\ \bibnamefont {Morgan}}, \bibinfo
  {author} {\bibfnamefont {R.}~\bibnamefont {Liang}}, \ and\ \bibinfo {author}
  {\bibfnamefont {K.}~\bibnamefont {Zhang}},\ }\href {\doibase
  10.1103/PhysRevLett.70.3999} {\bibfield  {journal} {\bibinfo  {journal}
  {Phys. Rev. Lett.}\ }\textbf {\bibinfo {volume} {70}},\ \bibinfo {pages}
  {3999} (\bibinfo {year} {1993})}\BibitemShut {NoStop}%
\bibitem [{\citenamefont {Lee-Hone}\ \emph {et~al.}(2017)\citenamefont
  {Lee-Hone}, \citenamefont {Dodge},\ and\ \citenamefont
  {Broun}}]{LeeHone2017}%
  \BibitemOpen
  \bibfield  {author} {\bibinfo {author} {\bibfnamefont {N.~R.}\ \bibnamefont
  {Lee-Hone}}, \bibinfo {author} {\bibfnamefont {J.~S.}\ \bibnamefont {Dodge}},
  \ and\ \bibinfo {author} {\bibfnamefont {D.~M.}\ \bibnamefont {Broun}},\
  }\href {\doibase 10.1103/PhysRevB.96.024501} {\bibfield  {journal} {\bibinfo
  {journal} {Phys. Rev. B}\ }\textbf {\bibinfo {volume} {96}},\ \bibinfo
  {pages} {024501} (\bibinfo {year} {2017})}\BibitemShut {NoStop}%
\bibitem [{\citenamefont {Zuev}\ \emph {et~al.}(2005)\citenamefont {Zuev},
  \citenamefont {Seog~Kim},\ and\ \citenamefont {Lemberger}}]{Zuev2005}%
  \BibitemOpen
  \bibfield  {author} {\bibinfo {author} {\bibfnamefont {Y.}~\bibnamefont
  {Zuev}}, \bibinfo {author} {\bibfnamefont {M.}~\bibnamefont {Seog~Kim}}, \
  and\ \bibinfo {author} {\bibfnamefont {T.~R.}\ \bibnamefont {Lemberger}},\
  }\href {\doibase 10.1103/PhysRevLett.95.137002} {\bibfield  {journal}
  {\bibinfo  {journal} {Phys. Rev. Lett.}\ }\textbf {\bibinfo {volume} {95}},\
  \bibinfo {pages} {137002} (\bibinfo {year} {2005})}\BibitemShut {NoStop}%
\bibitem [{\citenamefont {Benfatto}\ \emph {et~al.}(2008)\citenamefont
  {Benfatto}, \citenamefont {Castellani},\ and\ \citenamefont
  {Giamarchi}}]{Benfatto2008}%
  \BibitemOpen
  \bibfield  {author} {\bibinfo {author} {\bibfnamefont {L.}~\bibnamefont
  {Benfatto}}, \bibinfo {author} {\bibfnamefont {C.}~\bibnamefont
  {Castellani}}, \ and\ \bibinfo {author} {\bibfnamefont {T.}~\bibnamefont
  {Giamarchi}},\ }\href {\doibase 10.1103/PhysRevB.77.100506} {\bibfield
  {journal} {\bibinfo  {journal} {Phys. Rev. B}\ }\textbf {\bibinfo {volume}
  {77}},\ \bibinfo {pages} {100506} (\bibinfo {year} {2008})}\BibitemShut
  {NoStop}%
\bibitem [{\citenamefont {Yong}\ \emph {et~al.}(2013)\citenamefont {Yong},
  \citenamefont {Lemberger}, \citenamefont {Benfatto}, \citenamefont {Ilin},\
  and\ \citenamefont {Siegel}}]{Benfatto2013}%
  \BibitemOpen
  \bibfield  {author} {\bibinfo {author} {\bibfnamefont {J.}~\bibnamefont
  {Yong}}, \bibinfo {author} {\bibfnamefont {T.~R.}\ \bibnamefont {Lemberger}},
  \bibinfo {author} {\bibfnamefont {L.}~\bibnamefont {Benfatto}}, \bibinfo
  {author} {\bibfnamefont {K.}~\bibnamefont {Ilin}}, \ and\ \bibinfo {author}
  {\bibfnamefont {M.}~\bibnamefont {Siegel}},\ }\href {\doibase
  10.1103/PhysRevB.87.184505} {\bibfield  {journal} {\bibinfo  {journal} {Phys.
  Rev. B}\ }\textbf {\bibinfo {volume} {87}},\ \bibinfo {pages} {184505}
  (\bibinfo {year} {2013})}\BibitemShut {NoStop}%
\bibitem [{\citenamefont {Homes}\ \emph {et~al.}(2004)\citenamefont {Homes},
  \citenamefont {Dordevic}, \citenamefont {Strongin}, \citenamefont {Bonn},
  \citenamefont {Liang}, \citenamefont {Hardy}, \citenamefont {Komiya},
  \citenamefont {Ando}, \citenamefont {Yu}, \citenamefont {Kaneko} \emph
  {et~al.}}]{Homes2004}%
  \BibitemOpen
  \bibfield  {author} {\bibinfo {author} {\bibfnamefont {C.}~\bibnamefont
  {Homes}}, \bibinfo {author} {\bibfnamefont {S.}~\bibnamefont {Dordevic}},
  \bibinfo {author} {\bibfnamefont {M.}~\bibnamefont {Strongin}}, \bibinfo
  {author} {\bibfnamefont {D.}~\bibnamefont {Bonn}}, \bibinfo {author}
  {\bibfnamefont {R.}~\bibnamefont {Liang}}, \bibinfo {author} {\bibfnamefont
  {W.}~\bibnamefont {Hardy}}, \bibinfo {author} {\bibfnamefont
  {S.}~\bibnamefont {Komiya}}, \bibinfo {author} {\bibfnamefont
  {Y.}~\bibnamefont {Ando}}, \bibinfo {author} {\bibfnamefont {G.}~\bibnamefont
  {Yu}}, \bibinfo {author} {\bibfnamefont {N.}~\bibnamefont {Kaneko}},  \emph
  {et~al.},\ }\href {\doibase 10.1038/nature02673} {\bibfield  {journal}
  {\bibinfo  {journal} {Nature}\ }\textbf {\bibinfo {volume} {430}},\ \bibinfo
  {pages} {539} (\bibinfo {year} {2004})}\BibitemShut {NoStop}%
\bibitem [{\citenamefont {Varela}\ \emph {et~al.}(1999)\citenamefont {Varela},
  \citenamefont {Sefrioui}, \citenamefont {Arias}, \citenamefont {Navacerrada},
  \citenamefont {Luc\'{\i}a}, \citenamefont {L\'opez de~la Torre},
  \citenamefont {Le\'on}, \citenamefont {Loos}, \citenamefont
  {S\'anchez-Quesada},\ and\ \citenamefont {Santamar\'{\i}a}}]{Varela1999}%
  \BibitemOpen
  \bibfield  {author} {\bibinfo {author} {\bibfnamefont {M.}~\bibnamefont
  {Varela}}, \bibinfo {author} {\bibfnamefont {Z.}~\bibnamefont {Sefrioui}},
  \bibinfo {author} {\bibfnamefont {D.}~\bibnamefont {Arias}}, \bibinfo
  {author} {\bibfnamefont {M.~A.}\ \bibnamefont {Navacerrada}}, \bibinfo
  {author} {\bibfnamefont {M.}~\bibnamefont {Luc\'{\i}a}}, \bibinfo {author}
  {\bibfnamefont {M.~A.}\ \bibnamefont {L\'opez de~la Torre}}, \bibinfo
  {author} {\bibfnamefont {C.}~\bibnamefont {Le\'on}}, \bibinfo {author}
  {\bibfnamefont {G.~D.}\ \bibnamefont {Loos}}, \bibinfo {author}
  {\bibfnamefont {F.}~\bibnamefont {S\'anchez-Quesada}}, \ and\ \bibinfo
  {author} {\bibfnamefont {J.}~\bibnamefont {Santamar\'{\i}a}},\ }\href
  {\doibase 10.1103/PhysRevLett.83.3936} {\bibfield  {journal} {\bibinfo
  {journal} {Phys. Rev. Lett.}\ }\textbf {\bibinfo {volume} {83}},\ \bibinfo
  {pages} {3936} (\bibinfo {year} {1999})}\BibitemShut {NoStop}%
\bibitem [{\citenamefont {Chan}\ \emph {et~al.}(1993)\citenamefont {Chan},
  \citenamefont {Vier}, \citenamefont {Nakamura}, \citenamefont {Hasen},
  \citenamefont {Guimpel}, \citenamefont {Schultz},\ and\ \citenamefont
  {Schuller}}]{Chan1993}%
  \BibitemOpen
  \bibfield  {author} {\bibinfo {author} {\bibfnamefont {I.}~\bibnamefont
  {Chan}}, \bibinfo {author} {\bibfnamefont {D.}~\bibnamefont {Vier}}, \bibinfo
  {author} {\bibfnamefont {O.}~\bibnamefont {Nakamura}}, \bibinfo {author}
  {\bibfnamefont {J.}~\bibnamefont {Hasen}}, \bibinfo {author} {\bibfnamefont
  {J.}~\bibnamefont {Guimpel}}, \bibinfo {author} {\bibfnamefont
  {S.}~\bibnamefont {Schultz}}, \ and\ \bibinfo {author} {\bibfnamefont
  {I.~K.}\ \bibnamefont {Schuller}},\ }\href {\doibase
  https://doi.org/10.1016/0375-9601(93)90834-M} {\bibfield  {journal} {\bibinfo
   {journal} {Physics Letters A}\ }\textbf {\bibinfo {volume} {175}},\ \bibinfo
  {pages} {241 } (\bibinfo {year} {1993})}\BibitemShut {NoStop}%
\bibitem [{\citenamefont {Bluschke}\ \emph {et~al.}(2018)\citenamefont
  {Bluschke}, \citenamefont {Frano}, \citenamefont {Schierle}, \citenamefont
  {Putzky}, \citenamefont {Ghorbani}, \citenamefont {Ortiz}, \citenamefont
  {Suzuki}, \citenamefont {Christiani}, \citenamefont {Logvenov}, \citenamefont
  {Weschke} \emph {et~al.}}]{Bluschke2018}%
  \BibitemOpen
  \bibfield  {author} {\bibinfo {author} {\bibfnamefont {M.}~\bibnamefont
  {Bluschke}}, \bibinfo {author} {\bibfnamefont {A.}~\bibnamefont {Frano}},
  \bibinfo {author} {\bibfnamefont {E.}~\bibnamefont {Schierle}}, \bibinfo
  {author} {\bibfnamefont {D.}~\bibnamefont {Putzky}}, \bibinfo {author}
  {\bibfnamefont {F.}~\bibnamefont {Ghorbani}}, \bibinfo {author}
  {\bibfnamefont {R.}~\bibnamefont {Ortiz}}, \bibinfo {author} {\bibfnamefont
  {H.}~\bibnamefont {Suzuki}}, \bibinfo {author} {\bibfnamefont
  {G.}~\bibnamefont {Christiani}}, \bibinfo {author} {\bibfnamefont
  {G.}~\bibnamefont {Logvenov}}, \bibinfo {author} {\bibfnamefont
  {E.}~\bibnamefont {Weschke}},  \emph {et~al.},\ }\href {\doibase
  10.1038/s41467-018-05434-8} {\bibfield  {journal} {\bibinfo  {journal}
  {Nature communications}\ }\textbf {\bibinfo {volume} {9}},\ \bibinfo {pages}
  {1} (\bibinfo {year} {2018})}\BibitemShut {NoStop}%
\bibitem [{\citenamefont {Yang}\ \emph {et~al.}(2019)\citenamefont {Yang},
  \citenamefont {Liu}, \citenamefont {Wang}, \citenamefont {Feng},
  \citenamefont {He}, \citenamefont {Sun}, \citenamefont {Tang}, \citenamefont
  {Wu}, \citenamefont {Xiong}, \citenamefont {Zhang}, \citenamefont {Lin},
  \citenamefont {Yao}, \citenamefont {Liu}, \citenamefont {Fernandes},
  \citenamefont {Xu}, \citenamefont {Valles}, \citenamefont {Wang},\ and\
  \citenamefont {Li}}]{Yang2019}%
  \BibitemOpen
  \bibfield  {author} {\bibinfo {author} {\bibfnamefont {C.}~\bibnamefont
  {Yang}}, \bibinfo {author} {\bibfnamefont {Y.}~\bibnamefont {Liu}}, \bibinfo
  {author} {\bibfnamefont {Y.}~\bibnamefont {Wang}}, \bibinfo {author}
  {\bibfnamefont {L.}~\bibnamefont {Feng}}, \bibinfo {author} {\bibfnamefont
  {Q.}~\bibnamefont {He}}, \bibinfo {author} {\bibfnamefont {J.}~\bibnamefont
  {Sun}}, \bibinfo {author} {\bibfnamefont {Y.}~\bibnamefont {Tang}}, \bibinfo
  {author} {\bibfnamefont {C.}~\bibnamefont {Wu}}, \bibinfo {author}
  {\bibfnamefont {J.}~\bibnamefont {Xiong}}, \bibinfo {author} {\bibfnamefont
  {W.}~\bibnamefont {Zhang}}, \bibinfo {author} {\bibfnamefont
  {X.}~\bibnamefont {Lin}}, \bibinfo {author} {\bibfnamefont {H.}~\bibnamefont
  {Yao}}, \bibinfo {author} {\bibfnamefont {H.}~\bibnamefont {Liu}}, \bibinfo
  {author} {\bibfnamefont {G.}~\bibnamefont {Fernandes}}, \bibinfo {author}
  {\bibfnamefont {J.}~\bibnamefont {Xu}}, \bibinfo {author} {\bibfnamefont
  {J.~M.}\ \bibnamefont {Valles}}, \bibinfo {author} {\bibfnamefont
  {J.}~\bibnamefont {Wang}}, \ and\ \bibinfo {author} {\bibfnamefont
  {Y.}~\bibnamefont {Li}},\ }\href {\doibase 10.1126/science.aax5798}
  {\bibfield  {journal} {\bibinfo  {journal} {Science}\ }\textbf {\bibinfo
  {volume} {366}},\ \bibinfo {pages} {1505} (\bibinfo {year}
  {2019})}\BibitemShut {NoStop}%
\bibitem [{\citenamefont {Chiao}\ \emph {et~al.}(2000)\citenamefont {Chiao},
  \citenamefont {Hill}, \citenamefont {Lupien}, \citenamefont {Taillefer},
  \citenamefont {Lambert}, \citenamefont {Gagnon},\ and\ \citenamefont
  {Fournier}}]{Chiao2000}%
  \BibitemOpen
  \bibfield  {author} {\bibinfo {author} {\bibfnamefont {M.}~\bibnamefont
  {Chiao}}, \bibinfo {author} {\bibfnamefont {R.~W.}\ \bibnamefont {Hill}},
  \bibinfo {author} {\bibfnamefont {C.}~\bibnamefont {Lupien}}, \bibinfo
  {author} {\bibfnamefont {L.}~\bibnamefont {Taillefer}}, \bibinfo {author}
  {\bibfnamefont {P.}~\bibnamefont {Lambert}}, \bibinfo {author} {\bibfnamefont
  {R.}~\bibnamefont {Gagnon}}, \ and\ \bibinfo {author} {\bibfnamefont
  {P.}~\bibnamefont {Fournier}},\ }\href {\doibase 10.1103/PhysRevB.62.3554}
  {\bibfield  {journal} {\bibinfo  {journal} {Phys. Rev. B}\ }\textbf {\bibinfo
  {volume} {62}},\ \bibinfo {pages} {3554} (\bibinfo {year}
  {2000})}\BibitemShut {NoStop}%
\bibitem [{\citenamefont {Tomimoto}\ \emph {et~al.}(1999)\citenamefont
  {Tomimoto}, \citenamefont {Terasaki}, \citenamefont {Rykov}, \citenamefont
  {Mimura},\ and\ \citenamefont {Tajima}}]{Tomimoto1999}%
  \BibitemOpen
  \bibfield  {author} {\bibinfo {author} {\bibfnamefont {K.}~\bibnamefont
  {Tomimoto}}, \bibinfo {author} {\bibfnamefont {I.}~\bibnamefont {Terasaki}},
  \bibinfo {author} {\bibfnamefont {A.~I.}\ \bibnamefont {Rykov}}, \bibinfo
  {author} {\bibfnamefont {T.}~\bibnamefont {Mimura}}, \ and\ \bibinfo {author}
  {\bibfnamefont {S.}~\bibnamefont {Tajima}},\ }\href {\doibase
  10.1103/PhysRevB.60.114} {\bibfield  {journal} {\bibinfo  {journal} {Phys.
  Rev. B}\ }\textbf {\bibinfo {volume} {60}},\ \bibinfo {pages} {114} (\bibinfo
  {year} {1999})}\BibitemShut {NoStop}%
\bibitem [{\citenamefont {Rullier-Albenque}\ \emph {et~al.}(2003)\citenamefont
  {Rullier-Albenque}, \citenamefont {Alloul},\ and\ \citenamefont
  {Tourbot}}]{Albenque2003}%
  \BibitemOpen
  \bibfield  {author} {\bibinfo {author} {\bibfnamefont {F.}~\bibnamefont
  {Rullier-Albenque}}, \bibinfo {author} {\bibfnamefont {H.}~\bibnamefont
  {Alloul}}, \ and\ \bibinfo {author} {\bibfnamefont {R.}~\bibnamefont
  {Tourbot}},\ }\href {\doibase 10.1103/PhysRevLett.91.047001} {\bibfield
  {journal} {\bibinfo  {journal} {Phys. Rev. Lett.}\ }\textbf {\bibinfo
  {volume} {91}},\ \bibinfo {pages} {047001} (\bibinfo {year}
  {2003})}\BibitemShut {NoStop}%
\end{thebibliography}

%

\end{document}